\documentclass[conference]{IEEEtran}
\IEEEoverridecommandlockouts
\usepackage{cite}
\usepackage{amsmath,amssymb,amsfonts}
\usepackage{algorithmic}
\usepackage{graphicx}
\usepackage{textcomp}
\usepackage{xcolor}
\usepackage{amsfonts}
\usepackage{url}
\usepackage{diagbox}
\usepackage{multirow}
\usepackage{booktabs}

\def\BibTeX{{\rm B\kern-.05em{\sc i\kern-.025em b}\kern-.08em
    T\kern-.1667em\lower.7ex\hbox{E}\kern-.125emX}}
\begin{document}

\title{Deep Joint Source-Channel Coding for Efficient and Reliable Cross-Technology Communication \\
\thanks{This work is supported in part by The Major Key Project of PCL Department of Broadband Communication (PCL2023AS1-1), in part by the Science and Technology Development Fund, Macau SAR (File No. 0008/2022/AGJ), and in part by the China Postdoctoral Science Foundation under Grant Number 2023M741844.}
}

\author{
    \IEEEauthorblockN{Shumin Yao$^{1*}$, Xiaodong Xu$^{2,1}$, Hao Chen$^1$, Yaping Sun$^1$, and Qinglin Zhao$^{3}$}
    \IEEEauthorblockA{$^1$ Department of Broadband Communication, Peng Cheng Laboratory, Shenzhen 518066, China}
    \IEEEauthorblockA{$^2$ Beijing University of Posts and Telecommunications, Beijing 100876, China}
    \IEEEauthorblockA{$^3$ School of Computer Science and Engineering, Macau University of Science and Technology,\\ Avenida Wei Long, Taipa, Macau, China}
    \IEEEauthorblockA{yaoshm@pcl.ac.cn, xuxiaodong@bupt.edu.cn, \{chenh03, sunyp\}@pcl.ac.cn, qlzhao@must.edu.mo}
}
\maketitle

\begin{abstract}
Cross-technology communication (CTC) is a promising technique that enables direct communications among incompatible wireless technologies without needing hardware modification. However, it has not been widely adopted in real-world applications due to its inefficiency and unreliability. To address this issue, this paper proposes a deep joint source-channel coding (DJSCC) scheme to enable efficient and reliable CTC. The proposed scheme builds a neural-network-based encoder and decoder at the sender side and the receiver side, respectively, to achieve two critical tasks simultaneously: 1) compressing the messages to the point where only their essential semantic meanings are preserved; 2) ensuring the robustness of the semantic meanings when they are transmitted across incompatible technologies. The scheme incorporates existing CTC coding algorithms as domain knowledge to guide the encoder-decoder pair to learn the characteristics of CTC links better. Moreover, the scheme constructs shared semantic knowledge for the encoder and decoder, allowing semantic meanings to be converted into very few bits for cross-technology transmissions, thus further improving the efficiency of CTC. Extensive simulations verify that the proposed scheme can reduce the transmission overhead by up to 97.63\% and increase the structural similarity index measure by up to 734.78\%, compared with the state-of-the-art CTC scheme.
\end{abstract}


\section{Introduction} 

Cross-technology communication (CTC) enables direct communications among devices following incompatible wireless technologies such as WiFi, ZigBee, Bluetooth, etc., without hardware modification. This technique offers the potential to significantly simplify the massive heterogeneous connections in the 6G Internet of Things. 


However, CTC faces challenges in transmission efficiency and reliability due to the inherent hardware and standard mismatches between incompatible wireless technologies \cite{TwinBee}, hindering its widespread application in the industrial world. The inefficiency primarily arises from the mismatches between transmitting and receiving bandwidths. The maximum achievable rate in each cross-technology transmission is constrained by the technology with the narrowest bandwidth. Previous efforts to improve efficiency focused on designing coding schemes to enable parallel transmissions \cite{parallelWiFiZigBee, yao2020erfr}, neglecting to design source coding schemes to enhance efficiency per transmission. The unreliability can be attributed to the inherent mismatches between the signal transmission and reception standards. To tackle this problem, some studies incorporate channel coding schemes into their designs \cite{TwinBee, WEBee, NetCTC, longBee}. While these solutions are somewhat effective, they introduce redundancy, compromising the transmission efficiency. 

Ideally, efficient and reliable CTC can be achieved with both advanced source coding and channel coding schemes, where the former can reduce the redundancy introduced by the latter. However, designing two separate coding schemes is impractical for CTC because it can only achieve optimality with infinite block lengths and coding complexity, as proven by Shannon \cite{shannon1948mathematical}. Therefore, it would be better to adopt the joint design of both schemes, often referred to as joint source-channel coding, which is known to outperform the separation approach \cite{generativeDJSCC}. Nevertheless, practically implementing joint design has been a long-standing challenge. What is worse, there is no theoretical guideline for the joint design in CTC. This is mainly because of the persisting difficulty in mathematically modeling CTC links, which fundamentally differ from conventional wireless links \cite{zhang2020link}.

Recently, deep joint source-channel coding (DJSCC), a practical joint source-channel coding method based on the emerging deep learning, has shown outstanding results in achieving efficient and reliable communications \cite{generativeDJSCC}. In DJSCC, a sender employs a deep joint source-channel (DJSC) encoder to accomplish two critical tasks simultaneously: 1) intelligently compress messages to the point where only their essential semantic meanings are preserved; 2) precisely identify critical semantic meanings and provide protection to them. A receiver, in turn, adopts a DJSC decoder to accurately recover the message solely based on the received semantic meanings that are potentially corrupted. There have been studies on DJSCC. For example,  Xie et al. \cite{xie2021deep} designed a DJSCC scheme for text transmissions based on Transformer. Sun et al. \cite{zeroShutSKB} proposed a multi-level DJSCC powered by a semantic knowledge base for remote zero-shot object recognition. Wang et al. \cite{wang2022perceptual} proposed a perceptual learned DJSCC scheme for high-fidelity image semantic transmission.  Tung et al. \cite{tung2022deepjscc} designed a constellation-constrained DJSCC scheme, which is suitable for digital communication systems. Bo et al. \cite{bo2022learning} further explored this topic. However, all of these studies focused on in-technology communications (i.e., communications among devices with compatible technologies) rather than cross-technology communications.

We are inspired by the latest advancement of DJSCC to adopt this technique for efficient and reliable CTC. It allows a sender and a receiver who have incompatible technologies to communicate robustly with only the semantic meanings of their messages, by using a DJSC encoder-decoder pair. However, conventional DJSCC schemes are insufficient to achieve this goal, as they are designed only for in-technology communication. They do not account for the complex mismatches between heterogeneous hardware and standards, which are essential for finding the optimal encoding and decoding strategies for CTC. Hence, we need a DJSCC scheme that is specifically tailored for CTC.

This paper presents a pioneering DJSCC scheme that enhances the efficiency and reliability of CTC. The main contributions of this study are as follows. 
\begin{enumerate} 

\item We propose the first DJSCC scheme for efficient and reliable CTC. In the proposed scheme, the DJSC encoder and decoder are driven by two types of domain knowledge. First, they are driven by CTC knowledge to learn mismatches between heterogeneous hardware and standards, which helps them to find the best encoding and decoding strategies easily. Second, they are driven by semantic knowledge to convert messages into very few bits for cross-technology transmissions, thereby pushing the CTC efficiency to the extreme. 
\item We define and construct CTC and semantic knowledge for the proposed scheme. In addition, we design a shared knowledge base (KB) that allows CTC and semantic knowledge to be synchronized between the DJSC encoders and decoders.
\item We perform extensive simulations to validate the effectiveness of our DJSCC scheme in achieving efficient and reliable CTC. The simulation results show that our DJSCC scheme can reduce message transmission size by up to 97.63\% and improve the structural similarity index measure by up to 734.78\%, compared with the state-of-the-art CTC scheme. 

\end{enumerate} 

This study aims to improve the practicality of CTC.


The rest of the paper is organized as follows. Section \ref{sec:overview} explains the proposed DJSCC scheme. Section \ref{sec:training} presents the training challenges and the corresponding solutions. Section \ref{sec:experiments}  validates the proposed DJSCC scheme via simulatons. Finally, Section \ref{sec:conclusion} concludes this paper.

\begin{figure*}[h]\centering
    \centering
    \includegraphics[width=0.8\linewidth]{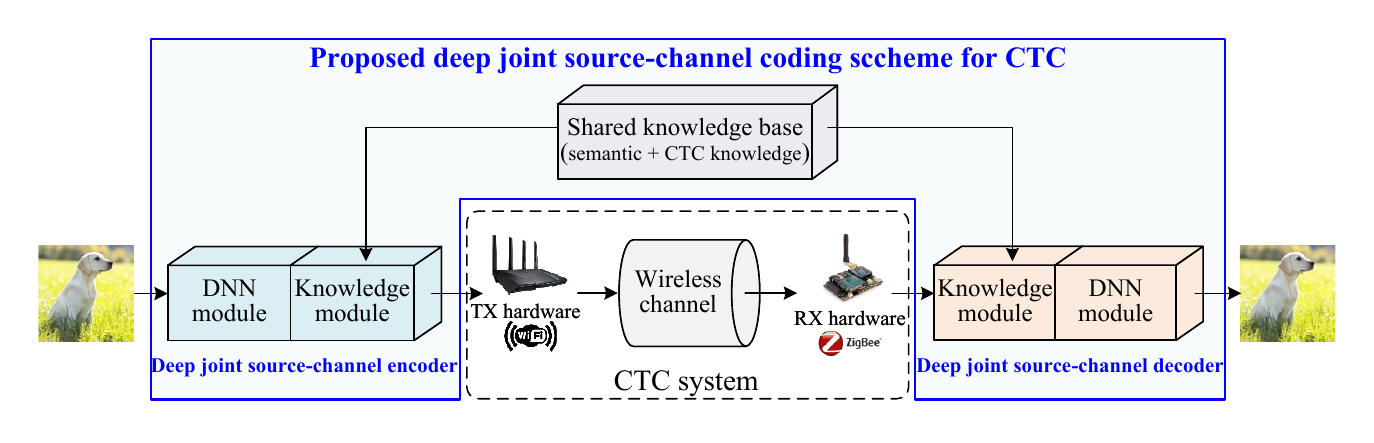}
    \caption{Framework of proposed DJSCC scheme.}
    \label{fig:framework}
\end{figure*}

\section{Proposed scheme} \label{sec:overview}

As shown in Figure \ref{fig:framework}, our DJSCC scheme extends a typical end-to-end CTC system, which consists of a sender and receiver adopting incompatible wireless technologies and a wireless channel between them. The scheme is designed based on the vector quantized variational autoencoder \cite{VQ-VAE} and consists of two components. The first component comprises a DJSC encoder and a DJSC decoder at the sender and receiver, respectively. The second component is a knowledge base (KB) shared between the DJSC encoder and decoder.

In each transmission, the sender employs the DJSC encoder, aided by the knowledge provided in the KB, to encode a message and then transmits the encoding result to the receiver through the CTC system. Upon receiving the encoding result, the receiver employs the DJSC decoder, with the help of the knowledge provided in the KB, to decode a message.

Below, we first design the components of the proposed DJSCC and then explain the encoding and decoding procedures.

\subsection{Component design}
In the following, we first explain the design of the DJSC encoder and decoder and then the design of the KB.

\subsubsection{DJSC encoder} \label{sec:encoder_design}

The DJSC encoder consists of a deep  neural network (DNN)  module followed by a knowledge module. The DNN module includes multiple blocks with convolutional, normalization \cite{Normalization}, and rectified linear unit activation layers. It also has some residual connections between layers for mitigating the vanishing gradient problem \cite{residualNet}. The knowledge module is responsible for pulling appropriate knowledge from the KB and incorporating it into the encoding process.

Overall, the encoder, driven by the knowledge from the KB, processes each input message by gradually reducing the spatial dimensions of the message while expanding a dimension relevant to its semantics. For example, if the message is an image, the encoder may decrease the image's height and width while increasing the image's channel size progressively. Ultimately, the encoder produces the semantic meanings extracted from the input message in a form that is robust to transmission through the CTC system.

\subsubsection{DJSC decoder} \label{sec:decoder_design}
The DJSC decoder consists of a knowledge module and a DNN module sequentially. The knowledge module contains a mechanism to pull appropriate knowledge from the KB and incorporate it into the decoding process. The DNN module can be roughly considered a mirror of the encoder's DNN module. 

The DJSC decoder acts as the inverse of the DJSC encoder. More specifically, driven by the knowledge from the KB, it reverses the encoding process to process the received potentially corrupted semantics meanings, gradually expanding the spatial dimensions and reducing the semantic dimensions of the vectors. Eventually, it reconstructs the source message inputted to the DJSC encoder.

\subsubsection{Knowledge base}
The KB comprises two kinds of knowledge: semantic knowledge and CTC knowledge. The semantic knowledge is defined as a set of $K$ semantic vectors, denoted as $\boldsymbol{\mathcal{S}} \triangleq \{\boldsymbol{s}_{1}, \dots, \boldsymbol{s}_{K}\} \in \mathbb{R}^{J \times K}$, where $s_{k}\in\mathbb{R}^{J}$, $1 \leq k \leq K$, is the $k$-th semantic vector. As will be explained in section \ref{sec:training}, the values of these vectors are learned via concurrent training with the deep modules of the DJSC encoder and decoder. As a result, a small set of vectors is sufficient to represent the semantic meanings of a message. This design allows for message reconstruction from these vectors, even in the presence of potential corruption, as if directly recovering the message from the semantic meanings. 

The CTC knowledge is defined as a set of $N$ existing CTC coding algorithms. These algorithms enable basic cross-technology transmissions of bits and have been developed over the years through extensive research regarding the mismatches between heterogeneous hardware and standards. Particularly, we denote the $n$-th algorithm, $n = 1, \dots, N$,  as a pair of functions $\mathcal{C}_{n} = \{C_{n}^{\text{TX}}(\cdot), C_{n}^{\text{RX}}(\cdot)\}$, where $C{n}^{\text{TX}}(\cdot)$ and $C_{n}^{\text{RX}}(\cdot)$ are the sender-side and the receiver-side operations of the algorithm, respectively. It is worth noting that if $\mathcal{C}_{n}$ is exclusively developed for the sender (receiver) side, $C_{n}^{\text{RX}}(\cdot)$ ($C_{n}^{\text{TX}}(\cdot)$) would simply be an identity function.

\subsection{Encoding and decoding procedures}
Assuming that both the sender and receiver have deployed $\mathcal{S}$ and $\mathcal{C}_{n}$, the encoding and decoding procedures are as follows.

\subsubsection{Encoding}
On the sender side, the DJSC encoder encodes a source message, denoted as $\mathbf{M}$, for transmissions through the following steps.

\underline{Step 1.} The encoder passes $\mathbf{M}$ through its DNN module, producing a set of vectors $\mathbf{X}=\{\boldsymbol{x}_{1},\dots,\boldsymbol{x}_{D}\} \in \mathbb{R}^{J \times D}$, with $\boldsymbol{x}_{d}\in\mathbb{R}^J$ and $1 \leq d \leq D$. That is, 
\begin{equation}\label{eq_encoding}
\mathbf{X} = F_{\alpha}(\mathbf{M}).
\end{equation}
Here, $F_{\alpha}(\cdot)$ represents the deep neural network used for constructing the DNN module in the DJSC encoder, where $\alpha$ represents a trainable network parameter.

\underline{Step 2.} The encoder exploits $\boldsymbol{\mathcal{S}}$ to fine-tune $\mathbf{X}$. More specifically, for $d=1,\dots,D$, its knowledge module first measures the distance between $\boldsymbol{x}_{d}$ and each semantic vector in $\boldsymbol{\mathcal{S}}$. Then, the knowledge module records the index, $z_d$, of the semantic vector that is the closest to $\boldsymbol{x}_d$:
\begin{equation}\label{vector2index}
    z_{d} = \mathop{\arg\min}\limits_{k = 1, \dots, K} 
    ||\boldsymbol{x}_{d}-\boldsymbol{s}_{k}||_{2},\forall \boldsymbol{s}_{k}, d=1,\dots , D.
\end{equation}

\underline{Step 3.} Let $\mathcal{Z}=\{z_{1},\dots,z_D\}$. The encoder exploits $\mathcal{C}_{n}$ to convert $\mathcal{Z}$ into a binary sequence, $\mathcal{B}_{1}$. More specifically, it adopts its knowledge module to perform the conversion using $C_{n}^{\text{TX}}(\cdot)$:
\begin{equation}
    \mathcal{B}_{1} = C_{n}^{\text{TX}}(\mathcal{Z}).
\end{equation}

After the encoding, the sender will follow the protocol of the wireless technology it adopts to pack $\mathcal{B}_{1}$ into packets and send them to the receiver through the CTC system directly. 

\subsubsection{Decoding}
The receiver follows the protocol of the wireless technology it adopts to receive the packets from the sender and extract a binary sequence, $\mathcal{B}_{2}$, from them. Here, it should be noted that $\mathcal{B}_{2} \neq \mathcal{B}_{1}$ due to the inherent mismatches between the transmission and reception standards. Then, based on $\mathcal{B}_{2}$, the receiver employs its DJSC decoder to reconstruct $\mathbf{M}$ through the following steps.

\underline{Step 1.} The DJSC decoder exploits $\mathcal{C}_{c}$ to convert $\mathcal{B}_{2}$ to $\hat{\mathcal{Z}}=\{\hat{z}_{1},\dots,\hat{z}_D\}$, which is the potentially corrupted version of $\mathcal{Z}$. More specifically, it adopts its knowledge module to perform the conversion using $C_{n}^{\text{RX}}(\cdot)$:
\begin{equation}
    \hat{\mathcal{Z}} = C_{n}^{\text{RX}}(\mathcal{B}_{2}).
\end{equation}

\underline{Step 2.} According to $\hat{\mathcal{Z}}$, the decoder adopts its knowledge module to retrieve the semantic vectors, $\hat{\mathbf{Q}}=\{\hat{\boldsymbol{q}}_{1},…,\hat{\boldsymbol{q}}_{D}\}\in\mathbb{R}^{J \times D}$, from $\boldsymbol{\mathcal{S}}$, where
\begin{equation}\label{index2vector}
    \hat{\boldsymbol{q}}_{d} = \boldsymbol{s}_{\hat{z}_{d}}.
\end{equation}
These retrieved vectors represent the semantic meanings of $\mathbf{M}$.

\underline{Step 3.} The receiver passes $\hat{\mathbf{Q}}$ through its DNN module to reconstruct $\mathbf{M}$. Let $\hat{\mathbf{M}}$ be the actual reconstruction result. We have
\begin{equation}
    \hat{\mathbf{M}} = F_{\beta}^{-1}(\hat{\mathbf{Q}}).
\end{equation} Here, $F_{\beta}^{-1}(\cdot)$ represents the deep neural network used for constructing the DNN module in the DJSC decoder, where $\beta$ represents a trainable network parameter. With proper $\alpha$ and $\beta$, we have $\hat{\mathbf{M}} = \mathbf{M}$.

It is important to acknowledge that the unreliable nature of the transmissions through the CTC system may result in the loss of some indices in $\hat{\mathcal{Z}}$. Without a complete set of $\hat{\mathcal{Z}}$, the receiver cannot obtain the completed $\hat{\mathbf{Q}}$ and, consequently, cannot obtain an accurate $\hat{\mathbf{M}}$. To address this issue, we replace the semantic vectors corresponding to the lost indices with the component-wise mean of semantic vectors in $\boldsymbol{\mathcal{S}}$. For example, if $\hat{z}_{d}$ is lost, its corresponding semantic vector, $\hat{\boldsymbol{q}}_{d}$, is replaced by
\begin{equation}
    \hat{\boldsymbol{q}}_{d}=\frac{\sum_{k=1}^{K}\boldsymbol{s}_k}{K}.
\end{equation}\\
\textit{Remarks:} Thanks to the sharing of $\boldsymbol{\mathcal{S}}$ between the DJSC encoder and decoder, only the indices of semantic vectors, which can be encoded with only a few bits, must be transmitted via the CTC system. This proves the effectiveness of incorporating shared semantic knowledge into DJSCC to improve CTC efficiency.

\addtolength{\topmargin}{0.01in}

\section{Training challenges and solutions} \label{sec:training}

To optimize our DJSCC scheme, we must jointly train the DJSC encoder and decoder along with an existing CTC system. However, this poses two challenges.

\textit{Challenge 1: Inconsistent gradient backpropagation.} During joint training, the gradients should be able to back-propagate from the DNN module in the DJSC encoder to the DNN module in the DJSC decoder without interruption. However, the CTC system and the knowledge modules between the deep modules interrupt the backpropagation because their operations are often non-differentiable. How to train without consistent gradient backpropagation poses a challenge.

\textit{Challenge 2: Construction of semantic vectors.} In our design, the semantic vectors play a crucial role in ensuring efficient and reliable cross-technology transmissions. For efficiency, we require a small set of semantic vectors to be sufficient to represent the semantic meanings of a message. For reliability, we require that the component-wise mean of the semantic vectors remains useful to the accurate message reconstruction when some vector indices are lost. Constructing such semantic vectors poses another significant challenge.

Inspired by vector quantised-variational autoencoder \cite{VQ-VAE}, we tackle the above challenges as follows. 

\textit{Solution to challenge 1.} We directly copy the gradients from the input of the DNN module in the DJSC decoder to the output of the DNN module in the DJSC encoder. Experiments validate the effectiveness of this approximation approach.

\textit{Solution to challenge 2.} We construct the semantic vectors by training them concurrently with the deep modules in the DJSC encoder and decoder. Initially, we randomly generate them. As the training progresses, we gradually optimize their values based on the message reconstruction performance. To this end, we design a loss function as follows:
\begin{equation}\label{loss}
    \begin{split}
        L(\hat{\mathbf{M}}, \mathbf{M}, \mathbf{X}, \hat{\mathbf{Q}}; \alpha, \beta) = ||\hat{\mathbf{M}} - \mathbf{M}||_{2} + \mu ||\mathrm{sg}[\mathbf{X}]-\hat{\mathbf{Q}}||_{2}\\ + \lambda ||\mathbf{X}-\mathrm{sg}[\hat{\mathbf{Q}}]||_{2}.
    \end{split}
\end{equation}

Note that \eqref{loss} consists of three components, each serving a crucial role in training:
\begin{itemize}
    \item The first component, $||\hat{\mathbf{M}} - \mathbf{M}||_{2}$, is a reconstruction loss, which measures the Euclidean distance between the original message $\mathbf{M}$ and its reconstruction result $\hat{\mathbf{M}}$. This component encourages the encoder and decoder to search for their optimal parameters (i.e., $\alpha$ and $\beta$) to achieve reliable message transmissions across incompatible wireless technologies.
    \item The second component, $\mu ||\mathrm{sg}[\mathbf{X}]-\hat{\mathbf{Q}}||_{2}$, is a semantic knowledge loss, which measures the Euclidean distance between the output, $\mathbf{X}$, of the DNN module in the DJSC encoder and the retrieved semantic vectors set $\hat{\mathbf{Q}}$ in the knowledge module in the DJSC decoder. Here, $\mu$ is a constant, and $\mathrm{sg}[\cdot]$ means no backward gradient propagation. This component encourages the semantic vectors in $\hat{\mathbf{Q}}$ to move toward $\mathbf{X}$, thereby capturing the semantic meanings.
    \item The third component, $\lambda ||\mathbf{X}-\mathrm{sg}[\hat{\mathbf{Q}}]||_{2}$, is a commitment loss, which also measures the Euclidean distance between $\mathbf{X}$ and $\hat{\mathbf{Q}}$, where $\lambda$ is another constant. However, different from the semantic knowledge loss, it ensures that $\mathbf{X}$ commits to $\hat{\mathbf{Q}}$. This component prevents the volumes of semantic vectors in SKB from growing uncontrollably, which could lead to instability during training.
\end{itemize}

\section{Performance evaluation} \label{sec:experiments}

In this section, we evaluate the performance of the proposed DJSCC scheme via simulation experiments. Our simulator consists of a DJSCC system and a CTC system, developed based on PyTorch 2.0 and MATLAB R2022b, respectively.

Our experiments focus on end-to-end WiFi-to-ZigBee image transmission. Therefore, the CTC system consists of WiFi nodes and ZigBee nodes, with the former transmitting images to the latter. The transmitted images are drawn from two popular datasets: CIFAR-10 and MNIST. Furthermore, an additive white Gaussian noise channel is assumed between two nodes.

To better demonstrate that our DJSCC scheme can effectively improve the efficiency and reliability of CTC, we compare it with WBee \cite{WEBee}, a well-known coding scheme enabling WiFi-to-ZigBee direct transmission. For a fair comparison, we assume that each transmission is carried by packets, each with a payload of 25 bytes, and there is no transmission feedback or repetition.

\begin{table}[h]
\centering
\caption{Network structure of DNN modules}
\label{tab:network}
\begin{tabular}{|l|l|}
\hline
\begin{tabular}[c]{@{}l@{}}DNN module \\ in DJSC encoder\end{tabular} &
  \begin{tabular}[c]{@{}l@{}}Conv(outCH=$D/2$, kernel=4, stride=2, padding=1)\\ BatchNormal\\ ReLU\\ Conv(outCH=$D$, kernel=4, stride=2, padding=1)\\ BatchNormal\\ ReLU\\ ResBlock(outCH=$D$)\\ BatchNormal\\ ResBlock(outCH=$D$)\\ BatchNormal\end{tabular} \\ \hline
\begin{tabular}[c]{@{}l@{}}DNN module \\ in DJSC decoder\end{tabular} &
  \begin{tabular}[c]{@{}l@{}}ResBlock(outCH=$D$)\\ BatchNormal\\ ResBlock(outCH=$D$)\\ BatchNormal\\ ConvTran(outCH=$D/2$, kernel=4, stride=2, padding=1)\\ BatchNormal\\ ReLU\\ ConvTran(outCH=$D$, kernel=4, stride=2, padding=1)\end{tabular} \\ \hline
ResBlock &
  \begin{tabular}[c]{@{}l@{}}ReLU\\ Conv(outCH=inCH, kernel=3, stride=1, padding=1)\\ ReLU\\ Conv(outCH=inCH, kernel=1, stride=1, padding=0)\end{tabular} \\ \hline
\end{tabular}
\end{table}

Regarding the settings of DJSCC, we choose WEBee as CTC knowledge in the knowledge modules and set $K=512$, $J=256$, $\mu = 0.5$, and $\lambda=0.25$. In addition, we let $D$ be 7 and 8 when transmitting images drawn from the MNIST and CIFAR-10 datasets, respectively. During training, we adopt the NAdam optimizer and evaluate the coding performance after only 40 epochs with a batch size 128. The network structure is listed Table \ref{tab:network}.

The figures below show the simulation results of both designs for the MNIST and CIFAR-10 datasets. For each dataset, we use two labels to denote the results of our design and the original WEBee design. For example, "Ours: MNIST" and "WEBee: MNIST" indicate the results of our design and that of the original WEBee design, respectively, when transmitting the images drawn from the MNIST dataset. Similarly, "Ours: CIFAR-10" and "WEBee: CIFAR-10" indicate the results of our design and that of the original WEBee design, respectively, when transmitting the images drawn from the CIFAR-10 dataset.

\begin{figure}[h]
    \centering
    \includegraphics[width=0.8\linewidth]{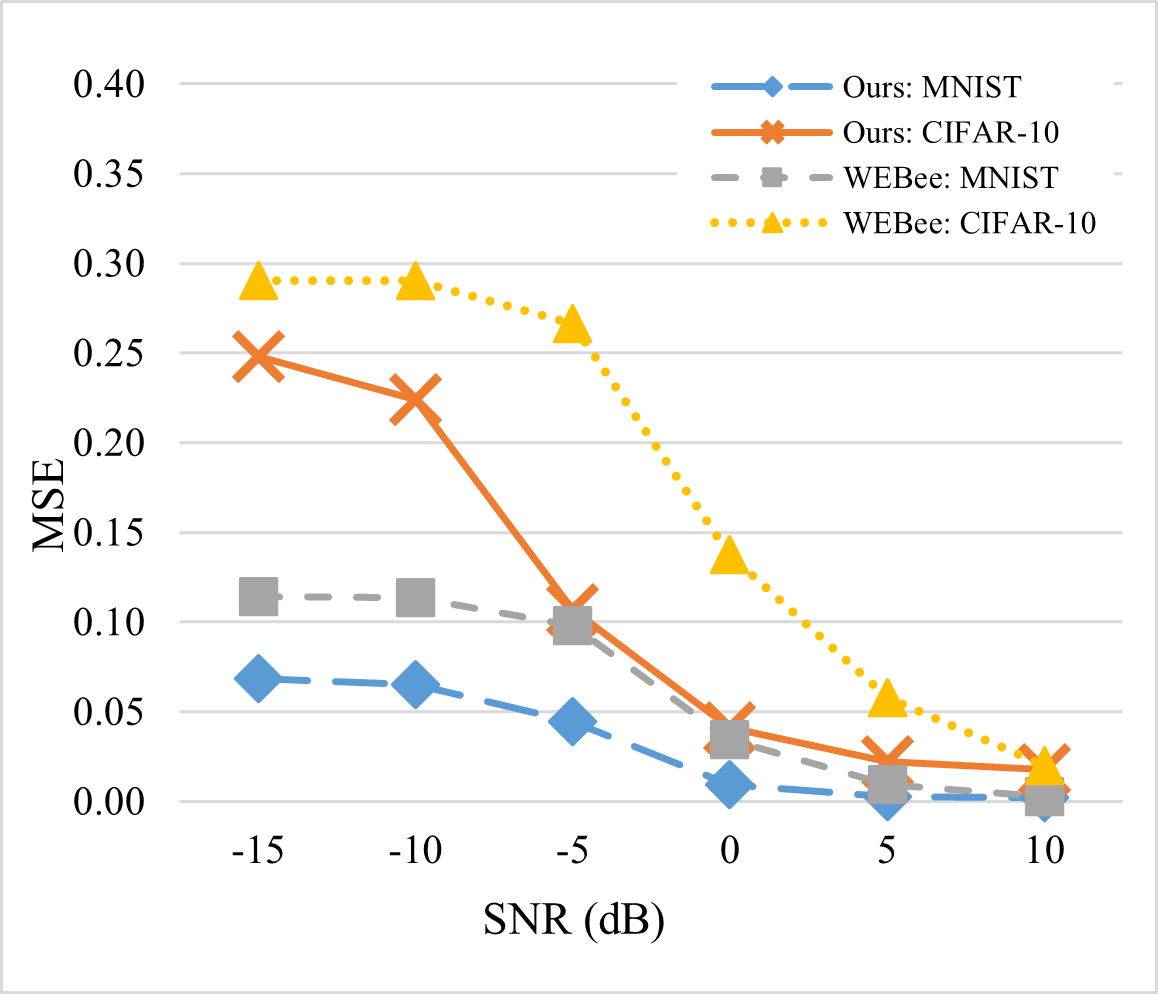}
    \caption{MSE under various SNR levels.}
    \label{fig:result_mse}
\end{figure}

Fig. \ref{fig:result_mse} plots the mean squared error (MSE) \cite{MSE} of the images on the receiver end (i.e., the ZigBee end) when the SNR varies. From this figure, we have the following observations.
\begin{enumerate}
    \item The MSEs of both our design and the original WEBee design decrease with increasing SNR. This is because a higher SNR leads to fewer incorrect receptions. In our design, fewer incorrect receptions allow the ZigBee node to obtain fewer incorrect indices of semantic vectors, reducing the image reconstruction inaccuracy. In the original WEBee, fewer incorrect receptions allow the ZigBee node to receive fewer corrupted pixels of the images transmitted by the WiFi node. 
    \item For images drawn from the same dataset, our design has a significantly lower MSE than the original WEBee design when the SNR is low (such as below 5 dB). This demonstrates the effectiveness of our DJSCC design in enhancing the reliability of CTC. Note that the SNR is consistently low in WiFi-to-ZigBee direct transmissions because ZigBee has a much narrower bandwidth than  WiFi and therefore can capture only a small fraction of the WiFi signal energy \cite{longBee}. 
    \item Given the same SNR, both designs have a lower MSE for MNIST images than for CIFAR-10 images. This is because MNIST images are grayscale images, which contain less information than the CIFAR-10 images that are colored. As a result, accurately receiving the MNIST images requires less effort than receiving the CIFAR-10 images. 
\end{enumerate}

\begin{figure}[h]
    \centering
    \includegraphics[width=0.8\linewidth]{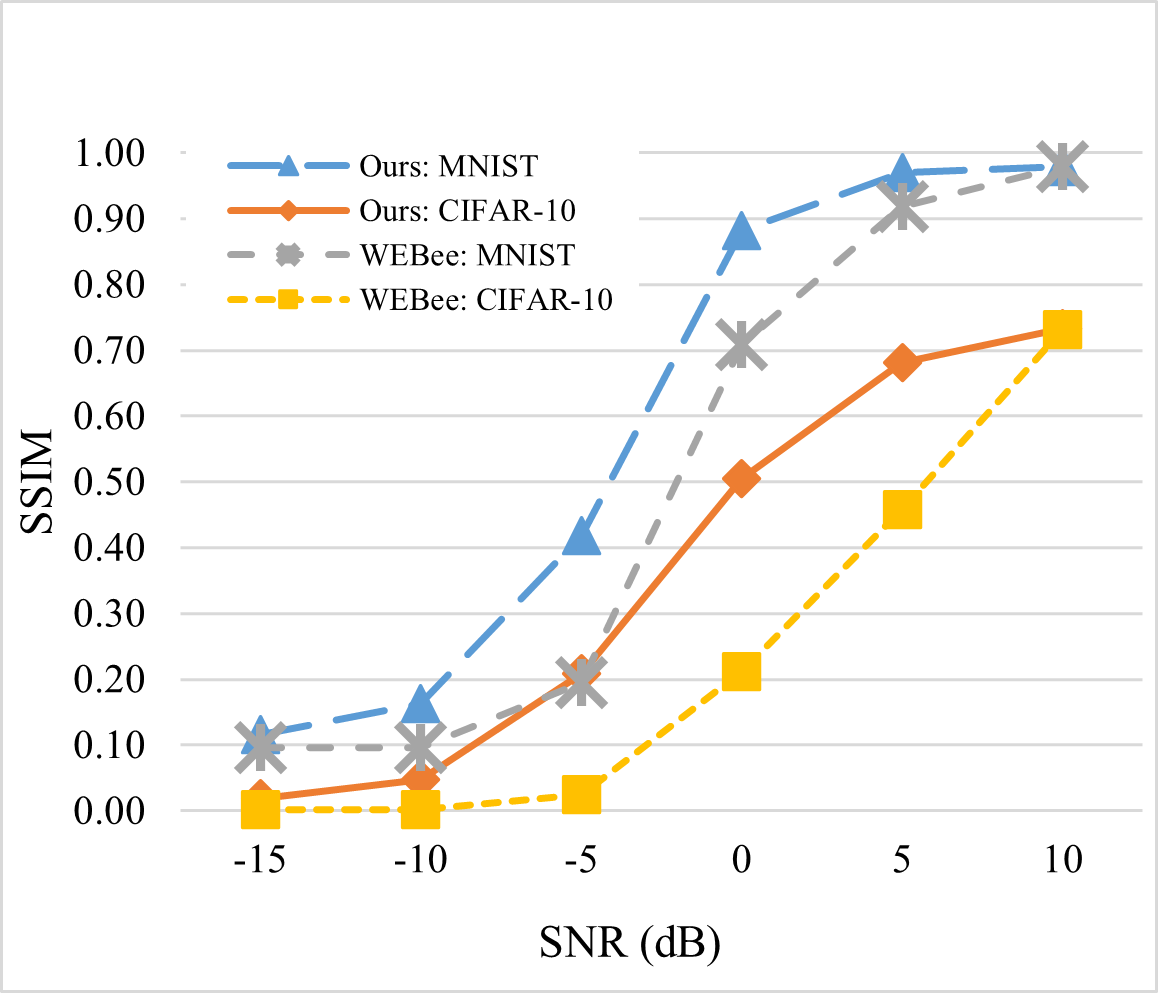}
    \caption{SSIM under various SNR levels.}
    \label{fig:result_ssim}
\end{figure}

Fig. \ref{fig:result_ssim} plots the structural similarity index measure (SSIM) between the transmitted and received images when the SNR$=-15,\dots,10$ dB. From this figure, we have the following observation.
\begin{enumerate}
    \item The SSIMs in both our design and the original WEBee design increase with SNR, as higher SNR leads to more correct receptions. In our design, more correct receptions allow the ZigBee to obtain more correct indices of semantic vectors, increasing the structural accuracy of image reconstructions. In the original WEBee design, the ZigBee node can receive images with higher structural accuracy from the WiFi node when the SNR is higher.
    \item For images drawn from the same dataset, our design has a significantly higher SSIM than the original WEBee design when the SNR is lower than 10 dB. Notably, our design outperforms the original by a remarkable 734.78\% when transmitting drawn from the CIFAR-10 dataset with an SNR of -5dB. This demonstrates the effectiveness of our DJSCC design for reliable CTC. 
    \item Given the same SNR, both designs have a higher SSIM for MNIST images than for CIFAR-10 images. This is because MNIST images are handwritten digits, which contain less structural information than the CIFAR-10 images showing ten different types of objects. As a result, accurately receiving the structural information of the former requires less effort than the latter's.
\end{enumerate}

Note that the proposed DJSCC design outperforms WEBee (shown in Fig. \ref{fig:result_mse} and Fig. \ref{fig:result_ssim}) with only 40 training epochs. This manifests the effectiveness of exploiting CTC knowledge in reducing the design and training difficulties of DJSCC for CTC.

\begin{figure}[h]
    \centering
    \includegraphics[width=0.8\linewidth]{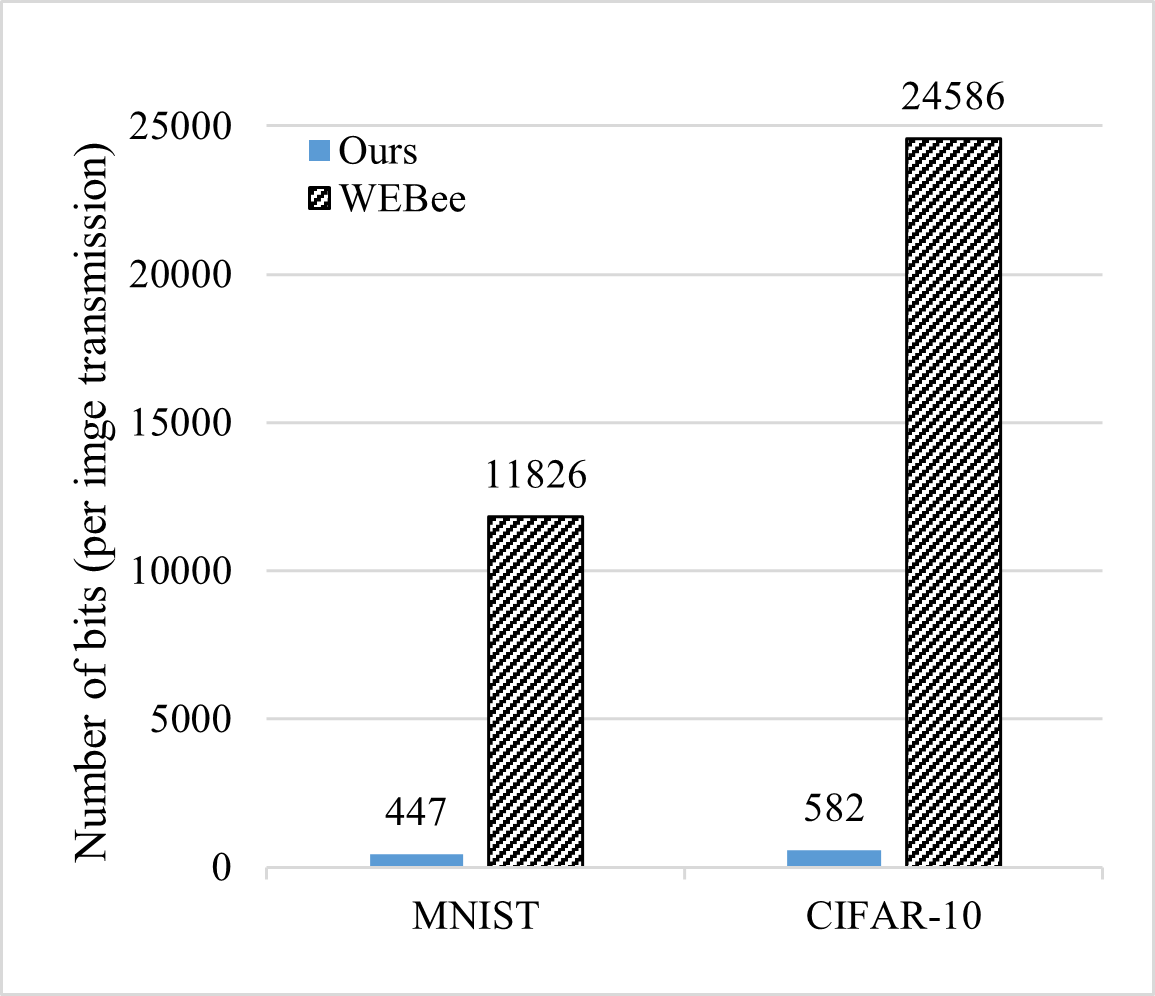}
    \caption{Number of bits per image transmission.}
    \label{fig:overhead}
\end{figure}

Fig. \ref{fig:overhead} shows the number of bits used for one image transmission when $K=512$. Here, it should be noted that in our design, bits represent the semantic vector indices and their coordinates. In WEBee, they represent image pixel values and coordinates. From the figure, we can see that our scheme uses significantly fewer bits per image transmission than WEBee, demonstrating the effectiveness of our design in enhancing CTC efficiency. In particular, for CIFAR-10 images, our design can reduce the number of bits per image transmission by 97.63\% compared to WEBee. The main reasons for this bit reduction are two-fold. First, our design enables the transmission of essential semantic meanings of images only. Second, the semantic-sharing structure of our design allows semantic meaning to be represented by very few bits for transmissions, further compressing the transmission amount. 

\section*{Conclusion} \label{sec:conclusion}
In this paper, we have proposed a DJSCC scheme to enable efficient and reliable CTC. Our scheme utilizes neural networks to compress and encode messages into semantic meanings and reliably transmit them across incompatible wireless technologies without needing hardware modification. Our scheme also adopts a semantic knowledge-sharing structure to reduce the transmission overhead. It incorporates existing CTC coding algorithms as knowledge to guide neural networks for better capturing the characteristics of the CTC links. We have conducted extensive simulations to evaluate the performance of our scheme and compared it with the state-of-the-art CTC scheme. The results show that our scheme can significantly improve the efficiency and accuracy of CTC, making it more suitable for real-world applications. We believe that our scheme opens up new possibilities for CTC and paves the way for future research in this area.

\bibliographystyle{ieeetr}
\bibliography{reference}

\end{document}